%% file: main_V1.tex
\documentclass[journal,10pt]{IEEEtran}

\usepackage{blindtext, graphicx}
\usepackage[english]{babel}
\usepackage[utf8]{inputenc}
\usepackage[T1]{fontenc}
\usepackage{multirow}
\usepackage[table]{xcolor}
\usepackage{graphics}

\usepackage{amsmath}
\usepackage{enumerate}
\usepackage{xcolor}
\usepackage{amsfonts}
\usepackage{mathtools}
\usepackage{amssymb}
\usepackage{mathrsfs}
\usepackage{epstopdf}
\usepackage{graphicx}
\usepackage{cite}
\usepackage{soulutf8}
\usepackage{ragged2e}

\usepackage{graphicx}
\usepackage{bm}
\usepackage{caption}

\usepackage{tabularray}

\usepackage{tablefootnote}
\usepackage[flushleft]{threeparttable}
\usepackage{pifont}
\usepackage{tikz}

\usepackage{subfigure}

\usepackage{mathtools, cuted}
\usepackage{lipsum, color}

\usepackage{xcolor}
\usepackage[hidelinks]{hyperref} 
\usepackage{academicons} 
\newcommand{\orcid}[1]{\href{https://orcid.org/#1}{\includegraphics[width=8pt]{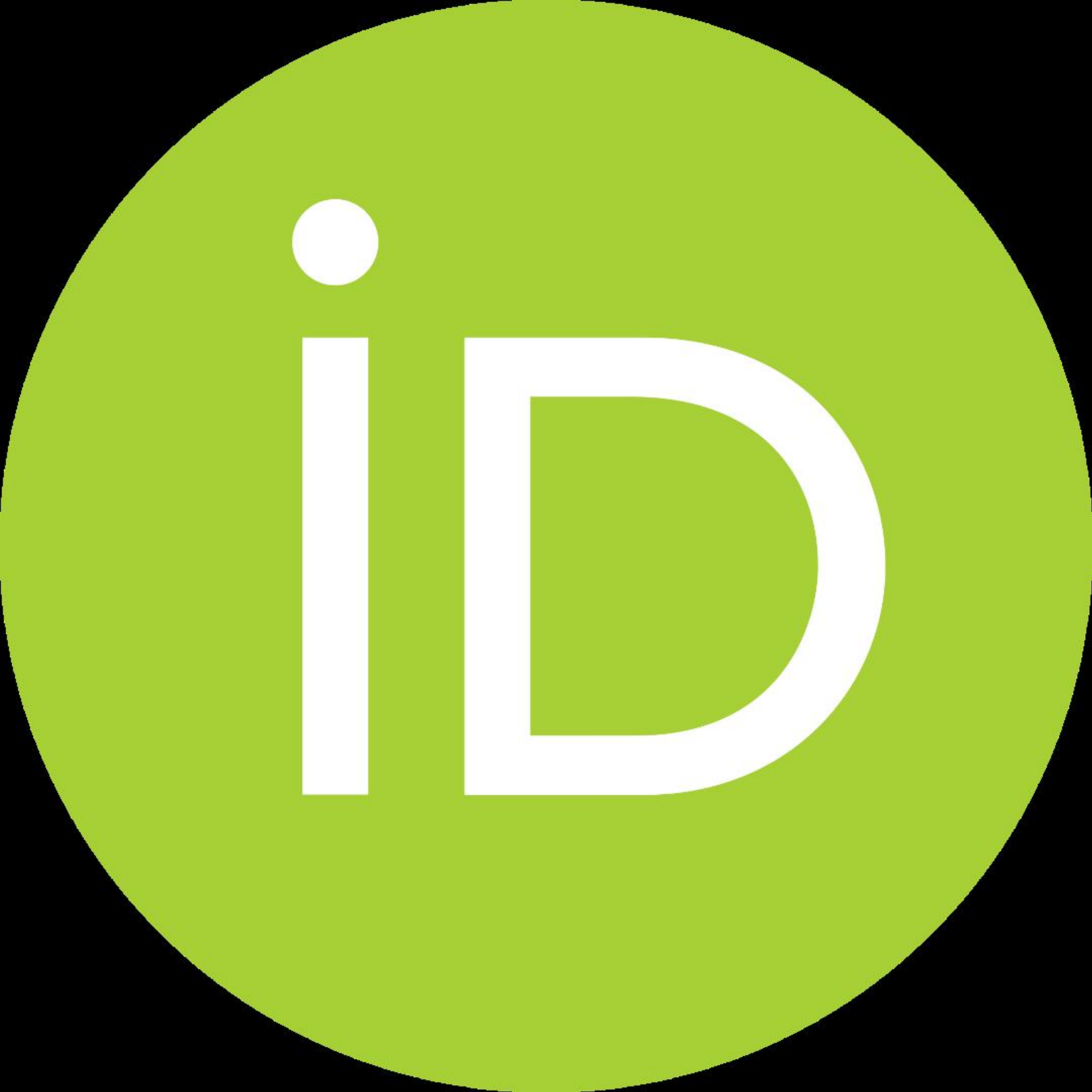}}}

\newcommand\felipe[1]{\textcolor{black}{#1}}
\newcommand\scd[1]{\textcolor{black}{#1}}

\begin{document}
\bstctlcite{IEEEexample:BSTcontrol}

\title{Non-Orthogonal  \felipe{Multiple-Access Strategies} for Direct-to-Satellite IoT Networks}
\author{{Felipe Augusto Tondo, Jean Michel de Souza Sant'Ana, \IEEEmembership{Member, IEEE}, Samuel Montejo-Sánchez, \IEEEmembership{Senior Member, IEEE}, Onel Luis Alcaraz López, \IEEEmembership{Member, IEEE}, Sandra Céspedes, \IEEEmembership{Senior Member, IEEE}, and \\ Richard Demo Souza, \IEEEmembership{Senior Member, IEEE}}
\thanks{F. A. Tondo and R. D. Souza are with the Federal University of Santa Catarina, Florianópolis, SC, Brazil. \{felipe.tondo@posgrad.ufsc.br, richard.demo@ufsc.br\}. S. Montejo-Sánchez is with the Instituto Universitario de Investigación y Desarrollo Tecnológico, Universidad Tecnológica Metropolitana, Santiago, Chile. \{smontejo@utem.cl\}. O. L. A. Lopez and J. M. S. Sant'Ana are with the Centre for Wireless Communications, University of Oulu, Finland. \{onel.alcarazlopez@oulu.fi, jean.desouzasantana@oulu.fi\}. S. Céspedes is with the Concordia University, Canada. \{sandra.cespedes@concordia.ca\}.}
\thanks{This work has been supported in Brazil by CNPq (402378/2021-0, 305021/2021-4, 401730/2022-0), Print CAPES-UFSC ``Automation 4.0'', and RNP/MCTIC (Grant 01245.010604/2020-14) 6G Mobile Communications Systems; in Chile by ANID FONDECYT Regular No. 1241977; in Finland by the Academy of Finland (6G Flagship Program, Grant 346208), The Finnish Foundation for Technology Promotion and by the European Union through the Interreg Aurora project ENSURE-6G (Grant Number: 20361812).}}


\maketitle

\begin{abstract}
    Direct-to-Satellite IoT (DtS-IoT) has the potential to support multiple verticals, including agriculture, industry, smart cities, and environmental disaster prevention. This work introduces two novel DtS-IoT schemes using power domain Non-Orthogonal Multiple Access (NOMA) in the uplink \felipe{with either fixed (FTP) or controlled (CTP) transmit power}. \felipe{W}e consider that the IoT devices use LoRa technology to transmit data packets to the satellite in orbit, equipped with a Successive Interference Cancellation (SIC)-enabled gateway. We also assume the IoT devices are empowered with a predictor of the satellite orbit. Using real geographic location and trajectory data, we evaluate the performance of the average number of successfully decoded transmissions, goodput (bytes/lap), and energy consumption (bytes/Joule) as a function of the number of network devices. 
    Numerical results show the trade-off between goodput and energy efficiency for both proposed schemes. Comparing FTP and CTP with regular ALOHA for $100$ ($600$) devices, we find goodput improvements of \scd{$65\%$ ($29\%$)} and \scd{$52\%$ ($101\%$)}, respectively. \felipe{Notably,} CTP effectively leverages transmission opportunities as the network size increases, outperforming the other strategies. Moreover, CTP shows the best performance in energy efficiency compared to FTP and ALOHA.    
\end{abstract}

\begin{IEEEkeywords}
Direct-to-Satellite, Internet-of-Things, LoRa, NOMA, SIC.
\end{IEEEkeywords}

\IEEEpeerreviewmaketitle

\section{Introduction}

\IEEEPARstart{C}{ontemporary} life dynamics \felipe{impose} an extraordinary challenge to upcoming wireless communication systems. From the inception of the first generation (1G) to the current era of the fifth generation (5G) wireless systems, researchers and industry have played important roles in providing global interconnection and raising \felipe{their} levels of performance and Quality of Service (QoS). However, a groundbreaking milestone was recently reached as the number of connected devices surpassed that of connected humans~\cite{IBM_EoT.23}. This notable advancement opens up a new opportunity for the so-called Economy of Things (EoT)~\cite{IBM_EoT.23}, providing a diversity of applications with nearly 30 billion connected Internet of Things (IoT) devices.

\felipe{Massive machine-type communication} (mMTC) is critical for the connectivity demands of smart cities, factories, and logistics, facilitating global connectivity~\cite{Choi_GF_NOMA.22}. \felipe{Still, remote} applications, like climate and maritime logistics, struggle with limited communication infrastructure, a gap filled by adopting non-terrestrial network (NTN) solutions and long-range IoT technologies~\cite{Hirley_DtS.23,Kodheli.21_survey_Sat5G}. The 3rd Generation Partnership Project (3GPP) aims at the upcoming sixth generation (6G) system in Release 20, focusing on the seamless integration of terrestrial and NTN solutions using advanced technologies to enable new applications like teleoperation, digital twins, and autonomous vehicles~\cite{Harounabadi.23}. Moreover, Direct-to-Satellite IoT (DtS-IoT) leads in NTN innovations by using gateways on satellites, reducing infrastructure costs but facing challenges like \felipe{highly} channel instability and multi-path losses \felipe{over the orbital trajectory}~\cite{Fraire.survey.22}. While geostationary satellites move with the Earth's orbit and provide fixed connectivity over an area, Low Earth Orbit (LEO) satellites move at around 7 km/s, serving different regions according to their movement. Although having more complicated dynamics, LEO satellites are much less expensive than geostationary satellites and are growing fast in number. 

Low power wide area networks (LPWAN) emerged as a notable alternative~\cite{Centenaro:CST:2021} among terrestrial technologies for wireless communication, with the potentiality to connect ground-based devices to satellites. The low power consumption, long-range distances, and minimal infrastructure requirements of LPWANs make them attractive for DtS-IoT applications. For instance, the LoRaWAN technology~\cite{LoRaWanParameters} for LPWANs employs LoRa modulation in the physical layer, while Lacuna Space is a pioneering company offering DtS-IoT LoRaWAN\felipe{. The latter} recently expanded its constellation coverage with the successful launch of Lacuna Space $2$nd-Gen, hosted on NanoAvionics modular satellite platform and launched on the Space-X Transporter$-7$~\cite{Lacuna_SpaceXF9.23}. 

Despite its simplicity, random access (RA) protocols, like those employed by LoRaWAN, tend to suffer from low scalability and high collisions in dense deployments~\cite{Georgiou.17}. In NTN, the scenario worsens due to dynamic factors and temporal visibility constraints imposed throughout the satellite coverage lap. Non-orthogonal multiple access (NOMA) approaches are candidate solutions to alleviate the number of unresolved collisions at the receiver and boost system efficiency~\cite{PD_NOMA_SAT.YAN.19}. Furthermore, many research efforts investigate the application of power domain NOMA in DtS-IoT scenarios, but they predominantly focus on the downlink~\cite{Gao.20_NOMA_Sat_Downlink, Basem.22_NOMA_Sat_Downlink}. Other studies improve the uplink performance using machine learning~\cite{Tubiana:WCL:2022} or cooperative non-orthogonal multiple access (C-NOMA)~\cite{Ge.22_NOMA_Sat_COOPERATIVE_UP} techniques. Regardless of the progress, the above approaches may be impractical due to the need of excessive gateway capabilities or extra information associated with device synchronization.

In this work, we propose novel uplink approaches for a LoRaWAN-based DtS-IoT network served by a LEO satellite, using power-domain NOMA while exploiting the knowledge of the satellite trajectory. We aim at improving the uplink scalability, without compromising the energy consumption and assuming no strict synchronization among devices.

\subsection{Related Works}

In \cite{Alvarez.22}, \felipe{the} authors present and analyze several LoRaWAN data rate optimization strategies for a DtS-IoT scenario. The devices know the satellite trajectory, and based on the channel state information, can \felipe{efficiently select the transmit} data rate. On top of that, several different approaches are proposed, including a centralized scheduling optimization based on a formal mixed integer linear programming model. \felipe{For this case}, the authors assume a central network server with knowledge on the device's traffic pattern and packet size. The authors in~\cite{F_Tondo.SJ.23} introduce a novel ALOHA-based traffic allocation strategy that achieves non-zero throughput even under high traffic loads \felipe{and} also increases the system performance in terms of energy efficiency. \felipe{However}, the method requires precise a priori knowledge of the traffic pattern.

The work in \cite{Alonso.23} proposes that the transmission time \felipe{of the devices} is randomized within the visibility window \felipe{ instead of allowing them to transmit as soon as there is a visible satellite}. Moreover, \felipe{the authors} propose adaptive schemes, where the devices choose to transmit or not based on \felipe{some} knowledge about the network traffic. Although it shows improved throughput and success probability, the algorithm relies on estimating the traffic load and the link success probability by using acknowledgments in the downlink. Even if the traffic does not vary due to changes in the network (devices entering and leaving), satellite laps generate different footprints with different devices in sight, considerably complicating practical deployment\felipe{s/implementations of} this approach.

\felipe{The au}thors in~\cite{Afhamisis:ACS:2022} propose SALSA, a time-division multiple access scheduling scheme for LoRa to LEO satellites.  It is assumed that the network server can estimate the satellite visibility windows for each device, whose locations are known, and then allocate a transmission slot, through a downlink communication, to selected devices. 
Moreover, to achieve higher capacity, the server needs to know the traffic pattern of \felipe{the} devices to avoid assigning too many slots to a device \felipe{without much data} to transmit. On a similar perspective, the work in \cite{Ortigueira.21} proposes a LoRa DtS-IoT access scheme where the satellite schedules transmissions for the devices. The whole procedure of satellite discovery and exchange of information is detailed. However, the request-to-transmit procedure may lead to overhead signaling, especially in cases with several devices requesting simultaneously. In the same line, the work in~\cite{F_Tondo.ASC.24} proposes two novel scheduling approaches for DtS-IoT with LoRa technology. The authors take advantage of multiple frequency channels to significantly increase uplink efficiency. However, \felipe{a} practical implementation requires the network server to know the device locations and traffic pattern\felipe{s}.


\felipe{Meanwhile}, the NOMA approach is advocated as a potential direction for \felipe{supporting} 6G ubiquitous IoT~\cite{Fang.21_NOMA_SAT_SURVEY}. An inherent issue in DtS-IoT is the high number of potential collided messages when many IoT devices transmit randomly in the uplink, \felipe{compromising performance parameters such as scalability and energy efficiency~\cite{Ortigueira.21}. Despite the efforts to address these problems in satellite-based IoT networks, the current literature does not explore non-orthogonal approaches using LoRaWAN protocols in DtS-IoT scenarios.} 

\begin{figure}[!t]
    \centering
    \includegraphics[width=0.9\columnwidth]{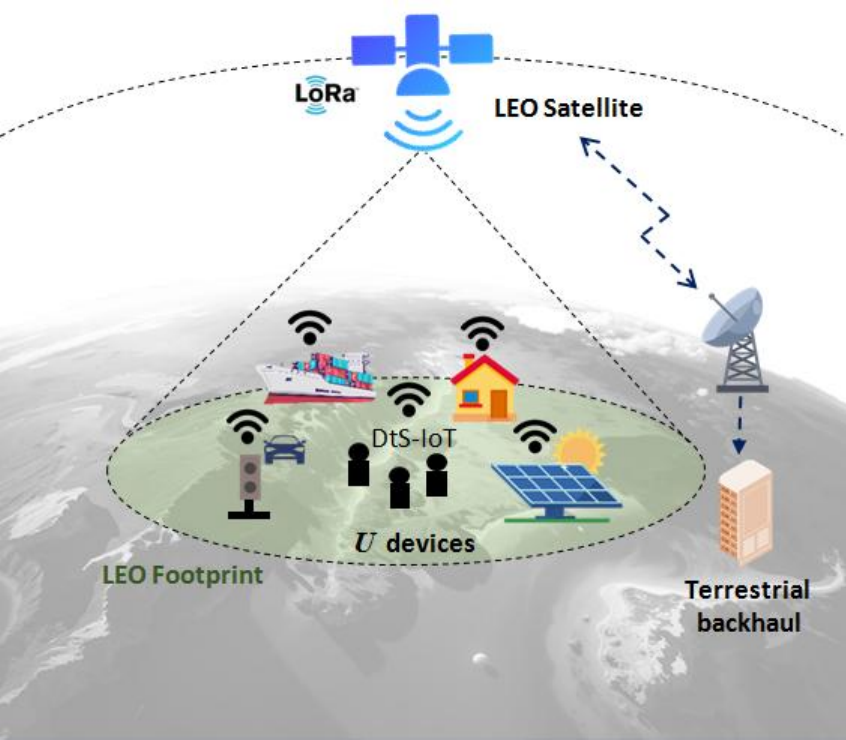} 
    \captionsetup{labelsep=period, name=Fig.}
    \caption{The DtS-IoT architecture consists of a gateway onboard a LEO satellite, IoT devices spread over the target area, and the terrestrial backhaul.} 
    \label{fig:System_Model}
\end{figure}

\subsection{Paper Contributions}

This work introduces two RA strategies for LoRa DtS-IoT, fixed transmission power (FTP) and controlled transmission power (CTP) methods. The strategies are designed such that transmissions from devices are received at the gateway with pre-defined average power levels. Such power levels are \felipe{set with} a separation such that two concurrent transmissions with different levels can be successfully decoded by SIC with high probability. In FTP, the devices choose when to transmit, within the visibility time window, to achieve one of the average power levels at the gateway. While in CTP, devices adapt their transmit power to transmit in different positions within the satellite visibility window, yielding one of the pre-determined average power levels at the satellite. Unlike~\cite{Alvarez.22, Alonso.23,Afhamisis:ACS:2022,F_Tondo.ASC.24}, the proposed schemes, FTP and CTP, are not dependent on network traffic estimation or immediate feedback links. Moreover, our methods are agnostic to the positioning of the devices, while solutions in \cite{Afhamisis:ACS:2022, F_Tondo.ASC.24,Alvarez.22} require such knowledge. Different from~\cite{Ortigueira.21}, we do not require any scheduling handshake procedure before transmission. {Numerical results show the trade-off between goodput and energy efficiency for both proposed NOMA-based schemes. \scd{Comparing with the regular ALOHA protocol for} $100$ ($600$) IoT devices over France, we find goodput improvements \scd{of $65\%$ ($29\%$) and $52\%$ ($101\%$)}.  Moreover, the CTP strategy is shown to be more energy efficient than FTP  and regular ALOHA.


Next, Section~\ref{sec:system_model} describes the system model, while Section~\ref{sec:dts-schemes} introduces the proposed DtS-IoT uplink strategies. Section~\ref{sec:results} presents the simulation parameters and discusses the numerical results. Finally, Section~\ref{sec:conclusion} concludes the paper and briefly discusses future works.

\section{System Model}
\label{sec:system_model}

We consider an IoT network composed of $U$ devices distributed within the target area and under the coverage of a single LEO satellite. We assume that the IoT devices use LoRa technology\footnote{\scd{The LoRa technology~\cite{semtech2015an1200} is based on CSS modulation. Recently, LR-FHSS modulation was added as an alternative in the LoRaWAN specification\cite{LoRaWanParameters}. There are interesting trade-offs in terms of performance, time-on-air and energy consumption related to CSS and LR-FHSS modulations~\cite{Semtech.AppNote}. Due to its widespread use and rich literature, in this work we focus on CSS LoRa modulation, but the extension of the proposed methods to LR-FHSS is relatively straightforward.}} to transmit data packets to the satellite in orbit, which is equipped with a SIC-enabled LoRa gateway, as seen in \felipe{Fig.}~\ref{fig:System_Model}. Meanwhile, the terrestrial backhaul consists of a ground station responsible for receiving the packets from the satellite and forwarding them to a network server (NS). \scd{For scope reasons, only the multiple access segment is of interest in this work.}

\begin{figure}[!t]
    \centering
    \includegraphics[width=1\columnwidth]{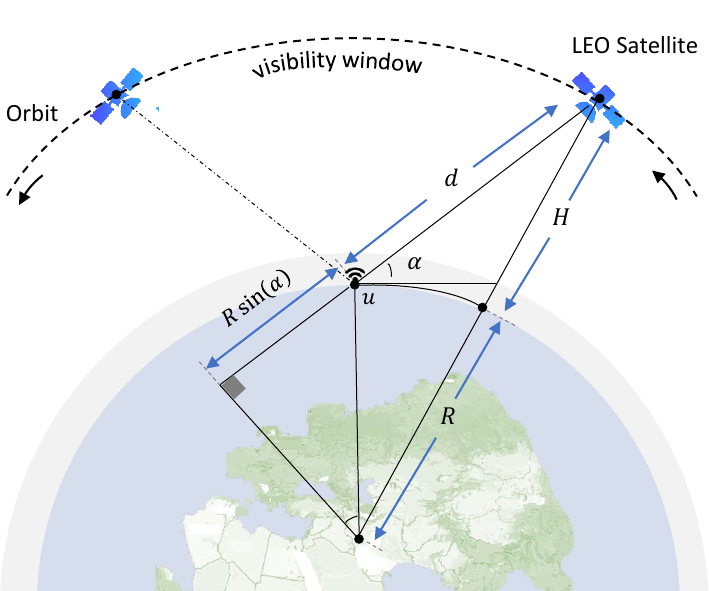} 
    \captionsetup{labelsep=period, name=Fig.}
    \caption{The ground-space geometry is described \scd{for the IoT device $u$} as a function of altitude $H$, distance $d$, elevation angle $\alpha$, and radius of Earth $R$.} 
    \label{fig:geometry}
\end{figure}

During the satellite visibility window, \textit{i.e.}, a visible orbit over the target area, also termed lap, each IoT device transmits a single message with a payload of $b$ bytes. We assume that the LEO satellite often and periodically broadcasts beacons to notify devices when they are under coverage. Suppose the \felipe{reference} IoT device $u\in \left \{ 1,2,...,U \right \}$ transmits a message, then the received signal at the satellite can be expressed as the sum of the attenuated transmitted signal, interference from the set $I$ \scd{containing} all other devices transmitting simultaneously with $u$, and noise, as
\begin{equation}
    r=\sqrt{P_u g_u} h_{u} s_{u} + \sum_{i\in I}  \sqrt{P_i g_i} h_{i} s_{i} + w,
\end{equation}
where, \felipe{for device $u$}, \felipe{$P_u$} is the transmit power, \felipe{$g_u$}  is the path loss, \felipe{$h_u$} is the channel fading and \felipe{$s_u$} is the modulated signal. Moreover, \felipe{$P_i$, $g_i$, $h_i$ and $s_i$ have the same meaning but for the $i$-th interfering device}, while $w$ is additive white Gaussian noise (AWGN) with zero mean and power $\sigma^{2}_w=-174+{F}+10\log_{10}B$ dBm, considering receive noise figure ${F}$ dB and bandwidth $B$ Hz~\cite{semtech2015an1200}. For the sake of clarity, in the following, the superscripts $u$ and $i$ are dropped whenever there is no ambiguity.

\begin{table}[!t]
\renewcommand{\arraystretch}{1.5} 
\centering
\caption{Non-terrestrial channel fading parameters $K$, $\mu$\felipe{,} and $\sigma$ as a function of elevation angle $\alpha$ \cite{Corazza.94}.}
\label{Tab:fading_parameters}
\begin{tabular}{c|c|c} 
\hline
\bm{$K(\alpha)$}        & \bm{$\mu(\alpha)$}           & \bm{$\sigma(\alpha)$ }   \\ \hline
$K_{0} + K_{1}\alpha+K_{2}\alpha^{2}$      & $\mu_{0} + \mu_{1}\alpha + \mu_{2}\alpha^{2}+ \mu_{3}\alpha^{3}$ & $\sigma_0+\sigma_1\alpha$     \\ \hline
\multicolumn{3}{c}{Coefficients for empirical formulas}         \\ \hline
\multirow{4}{*}{\begin{tabular}[c]{@{}c@{}}$K_0=2.731$ \\$K_1=-1.0474 \ 10^{-1}$ \\$K_2-2.7740 \ 10^{-3}$\end{tabular}} & $\mu_0=-2.331$    & \multirow{4}{*}{\begin{tabular}[c]{@{}c@{}}$\sigma_0=4.5$ \\$\sigma_1=-0.05$\end{tabular}}  \\   & $\mu_1=1.142 \ 10^{-1}$    &      \\           & $\mu_2=-1.939 \ 10^{-3}$            &  \\ & $\mu_3=-1.094 \ 10^{-5}$            &              \\
\hline
\end{tabular}
\end{table}

The visibility window of each device occurs while the elevation angle between the satellite and that device is above a threshold $\alpha_{\min}$. The visibility window consists of two phases: first, the ascendant phase, where the elevation angle goes from minimum to maximum ($\alpha_{\min} \rightarrow \alpha_{\max}$), while in the descendant phase, it goes from maximum to minimum ($\alpha_{\max} \rightarrow \alpha_{\min}$). Moreover, as devices experience different maximum elevation angles, they also experience distinct visibility windows given a satellite lap. The duration of a visibility window for a given device is determined by the interval 
 limited by the rise time and set time. The rise time happens when $\alpha = \alpha_{\min}$ during the ascendant phase, while the set time occurs \scd{again when $\alpha = \alpha_{\min}$, but} during the descendant phase. 

In \felipe{Fig.~}\ref{fig:geometry}, we illustrate a DtS-IoT link, showing the \scd{elevation} angle and the parameters required to calculate the distance between a device \scd{$u$} and the satellite. The distance is a function of the elevation angle $\alpha$~\cite{Hirley_DtS.23,Maleki.23}
\begin{align}
    d(\alpha)&=\left [ \sqrt{\left ( R\!+\!H \right )^{2}-(R \cos(\alpha ))^{2}} - R \sin(\alpha ) \right ], \label{eqn:d_a}
    \end{align}
where $\alpha$ is given in degrees, $R=6.378\times10^6$~m is  the Earth radius and $H$ is the orbital height of the LEO satellite.

We model the path loss using the {free-space formula~\cite{rappaport2024wireless}} 
\begin{equation}
    g = G_{t}G_{r} \left ( \frac{\lambda}{4\pi d} \right )^2, \label{eqn:pathloss}
\end{equation}where $\lambda=c/f$ is the wavelength, $c$ is the speed of light,  $f$ is the carrier frequency, and $G_{t}$ and $G_{r}$ are the antenna gains at the transmitter and receiver, respectively. 


\subsection{Non-Terrestrial  Fading Model}

Due to the non-geostationary orbit, the elevation angle $\alpha$ changes with time, modifying the relative channel conditions between a device and the satellite. There {are many available models in the literature~\cite{Loo.85,Fontan_Loo2001,Abdi.Nakagami.2003,Akturan_Nakagami.95},} such as Loo and Nakagami~-~m distributions that can be used to characterize the fading in such setups, while works as~\cite{Choi_GF_NOMA.22,Corazza.94, Salamanca.19} advocate that Rice fading is an attractive approach for ground-to-space links. Inspired by \scd{\cite{Corazza.94}}, we model the fading envelope $h$ as a combination of two processes $f_h(h|\mathcal{S})$ and $f_{\mathcal{S}}(\mathcal{S})$: 
\begin{equation}
    f_h(h)=\int_{0}^{\infty} f_h(h|\mathcal{S}) f_{\mathcal{S}}(\mathcal{S}) \scd{.}
\label{eq:overall_RxL}
\end{equation}
First, $f_h(h|\mathcal{S})$ is the Rice probability distribution function (PDF) parameterized according to the shadowing $\mathcal{S}$, as~\cite{Corazza.94}:
\begin{align}
    f_h(h|\mathcal{S}) &= 2(K+1)\frac{h}{\mathcal{S}^{2}}  \mathrm{exp} \left[- (K+1)\frac{h^{2}}{\mathcal{S}^{2}}-K \right]  \nonumber \\
    &\times \ \ I_o\left(2 \frac{h}{\mathcal{S}}\sqrt{K(K+1)} \right),
\end{align}
where $I_o$ {is the zero-order modified Bessel function~\cite{Bell.2004}} while $K$ is the Rice factor, the ratio between the power in the line-of-sight (LOS) component over the non-line-of-sight components (NLOS).  Moreover, we model the log-normal shadowing $S$ as~\cite{Corazza.94}:
\begin{equation}
    f_{\mathcal{S}}(\mathcal{S}) = \frac{1}{\sqrt{2\pi} \beta \sigma \mathcal{S}} \mathrm{exp}\left [ -\frac{1}{2} \left ( \frac{\mathrm{ln} \ S - \mu}{\beta \sigma} \right )^{2} \right ],
\end{equation}
where $\beta=(\mathrm{ln} 10)/20$, $\mu$ and $(\beta \sigma)^2$ are the mean and variance of the associated normal variate.  In Table~\ref{Tab:fading_parameters}, we list parameters $K$, $\sigma$, and $\mu$ as a function of the elevation angle for a rural tree-shadowed environment~\cite{Corazza.94}. \felipe{ Herein, a} larger $\alpha$ results in smaller $\sigma$ and larger Rice factor $K$, \textit{i.e.}, increased channel LOS and better link conditions. Note that the model in~\cite{Corazza.94} is based on actual measurements.

\subsection{Conditions for Successful Decoding}
\label{sec:sic}

Following the literature on LoRa networks~\cite{Aamir.19, Alvarez.22, Alonso.23}, we assume two conditions must be met to guarantee successful decoding at the gateway. The first condition is that the signal to noise ratio (SNR) at the gateway must be above a threshold $\gamma$ defined by the LoRa technology~\cite{semtech2015an1200}, which is a function of the spreading factor (SF). Therefore, the received power must be high enough for the gateway to detect and decode the message over the noise level. The SNR at the gateway given a transmission of  device $u$ can be written as
\begin{align}
    \text{SNR}_u = \frac{P_ug_u h_u^2 }{\sigma_{w}^2}, 
\end{align}so that the first condition for successful decoding is 
\begin{align}
    {\cal C}1=  \text{SNR}_u \geq \gamma.
    \label{Eq:C1}
\end{align}

The second condition states that the signal to interference ratio (SIR) at the receiver must be above the capture threshold $\psi$, which is fixed for a given technology and it is well studied in the LoRa case~\cite{Croce_CL_2018}. The received power of a given transmission must be sufficiently higher than the sum of the interference (\textit{i.e.}, other transmissions happening at the same time). We express the SIR of a given transmission as
\begin{align}
    \text{SIR}_u = \frac{P_u g_u h_u^2 }{\sum\limits_{i \in I} P_i g_i h_i^2}.
\end{align}Therefore, the second condition for successful decoding is
\begin{align}
    {\cal C}2=  \text{SIR}_u \geq \psi.
    \label{Eq:C2}
\end{align}


\felipe{In this work,} we assume that \felipe{once a collision between transmitted messages happens,} the gateway may be able to apply SIC and potentially recover them. \felipe{For} that sake, besides meeting conditions ${\cal C}1$ and ${\cal C}$2, the messages must have been transmitted using orthogonal pilots (or syncwords). This additional condition is to guarantee that the gateway can, after decoding the stronger message, estimate the channel concerning the stronger user, correctly reproduce what would be the received signal corresponding to that transmitted message, remove its contribution from the overall received signal (containing the collision of all messages), and then decode the message of the second strongest user from the remaining signal. This process iterates until the weakest user is decoded.

\begin{figure}[!t]
    \centering
    \includegraphics[width=1\columnwidth]{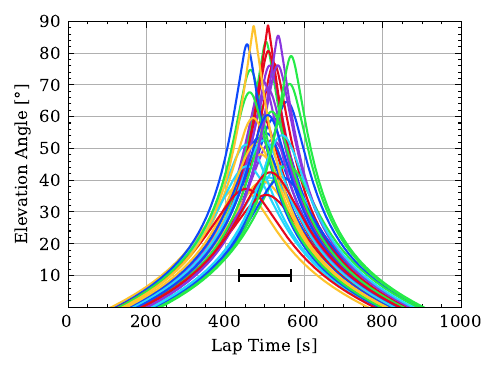} 
    \captionsetup{labelsep=period, name=Fig.}
    \caption{The perceived elevation angle, with respect to the satellite, {for $U=40$} end-devices randomly deployed over the territory of France for an actual satellite lap (LacunaSat-3). The black line represents the time from the first to last maximum elevation angle of the devices.} 
    \label{fig:DtS_Pr}
\end{figure}

\section{Proposed Schemes}
\label{sec:dts-schemes}

In this section, we introduce two novel uplink strategies for DtS-IoT. Moreover, we assume the IoT devices are empowered with a predictor of the satellite orbit. Given this knowledge, a straightforward strategy would be to use plain ALOHA and let devices transmit \felipe{freely} in their visibility windows\felipe{. This} is very simple to implement but may lead to a large number of potential colliding devices, as several of them may have non-zero intersection in their visibility windows. Another possibility would be to let devices transmit in a given particular position, \felipe{such as} at the time instant corresponding to their maximum elevation angle. This sounds reasonable, as this position corresponds to the minimum distance concerning the satellite, leading to a large received power. However, this strategy has the side effect of reducing the overall effective visibility window of the satellite, increasing the collision probability. This is illustrated in~\felipe{Fig.~}\ref{fig:DtS_Pr}, where we consider a real study case of the LacunaSat-3~\cite{LS3.2020} satellite orbiting over the region of France, with a total visibility window of around 800 seconds. The curves show the elevation angle versus time for each device, while the black line shows the  interval from the time the first device reaches its maximum elevation angle to the time the last device reaches its maximum elevation angle, a period that comprises only 180 seconds, thus reducing the overall satellite visibility window in more than four times. If devices transmit at the maximum elevation angle only, then the period represented by the black line would contain all the transmissions, increasing the perceived traffic load by the satellite on that period, and also increasing the collision probability. 

\felipe{A} promising alternative may be one that lays between plain ALOHA and a strategy with a single transmit opportunity per device. An option is to let devices transmit in some positions spread within their visibility windows, so that due to their different geographical location it would be unlike to have a collision. Moreover, another interesting possibility is to consider power domain NOMA, while employing SIC at the gateway\felipe{. In such a case,} more than one device may successfully transmit at the same time instant, \felipe{alleviating} the issue with the plain ALOHA strategy. Next, we present two novel uplink strategies that exploit the above ideas.

\subsection{Fixed Transmit Power (FTP)}

In the FTP strategy, the devices employ a fixed transmit power, while they  choose the transmit position\footnote{\scd{We refer to the transmit position within the trajectory of the LEO satellite, such as the initial position (rise time) and the end position (set time).}} within their visibility windows so that the average received power at the gateway is $\mathcal{L}_l$, for $l \in \{1, 2, \ldots, L\}$, where $L$ is the number of predefined received power levels. This is accomplished by inverting \eqref{eqn:pathloss}, considering a fixed transmit power $P$, obtaining
\begin{align}
    d^{\text{FTP}} = \frac{\lambda \sqrt{PG_tG_r}}{4\pi\sqrt{\mathcal{L}_l}}. \label{eqn:ftp}
\end{align}Thus, with a fixed transmit power $P$, devices choose the appropriate time to transmit such that their distance to the satellite $d^{\text{FTP}}$ results in the desired path loss, and consequently, the desired target received power level. Note that, due to the symmetric nature of the satellite orbit over a given region, there is a maximum of $2 \times L$ possible transmit positions for a device: $L$ times during the ascending phase and $L$ times during the descending phase. The utilized transmit position should be randomly chosen by the device within the possible positions. Furthermore, we assume that the power levels have sufficient  difference among them so that applying NOMA/SIC is possible, and that each power level is associated with an orthogonal pilot, for the reasons mentioned in \felipe{Section~\ref{sec:sic}}. 

Moreover, note that there is a relation between the distance and the elevation angle, as given by \eqref{eqn:d_a}. For orbits with low maximum elevation angle, a device may not be able to achieve certain power levels, as they would required a high elevation angle (reduced path loss) that is just not possible in that particular satellite lap for that device. Finally, note that transmissions are subject to fading and shadowing, causing the received power to deviate from the target power levels, so that SIC may fail if the actual received power levels are not sufficiently apart.

\subsection{Controlled Transmit Power (CTP)}

The FTP strategy limits the transmissions to particular slots within the visibility windows of each device, thus contributing to reduce collisions. However, it is vulnerable to a potential contender device with a very similar distance to the satellite. To address this limitation, the CTP strategy allows devices to transmit at any time during the orbit by adapting their transmit power, while respecting a maximum transmit power constraint, therefore spreading more their transmissions within the visibility window while still employing NOMA. In this scenario, by  \felipe{rearranging} \eqref{eqn:ftp} we find the required transmit power to yield a given average received power at the satellite considering a particular position within the visibility time window as
\begin{align}
    P^{\text{CTP}} = \frac{16\pi^2 d^2 \mathcal{L}_l}{\lambda^2 G_t G_r}. \label{eqn:ctp}
\end{align}

\subsection{Performance Metrics} 

The performance of the proposed methods is evaluated in terms of two metrics, the goodput  $\mathcal{G}$ and the energy efficiency  $\mathcal{E}$.
The goodput is defined as the average number of successfully received bytes at the satellite per lap (bytes/lap)
\begin{equation}
    \mathcal{G} = \mathcal{M} b,
    \label{eq:goodput}
\end{equation}
considering the number of successfully received messages \felipe{$\mathcal{M}\leq U$} at the satellite during that lap and the number of bytes $b$ per message.

The energy efficiency, in bytes/J, is the ratio between the goodput and the average transmit power ${\bar  P}$ used by the end devices during that lap, so that it can be calculated as
\begin{equation}
    \mathcal{E} = \frac{\mathcal{G}}{U{\bar  P}}.  \label{eq:energy_efficiency}
\end{equation}


\begin{figure*}[!t]
     \hspace{-0.4cm}
     \begin{subfigure}{}
         \centering
         \includegraphics[width=1\columnwidth]{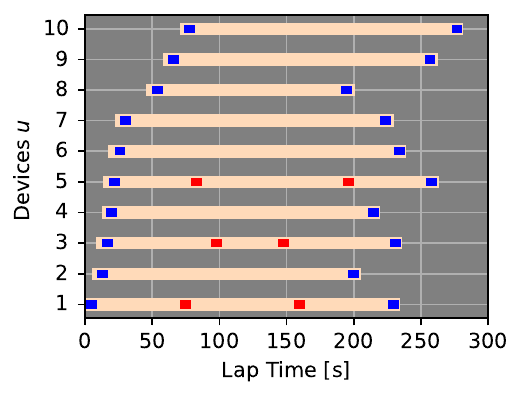}
         \hspace{0.3cm}
         \includegraphics[width=1\columnwidth]{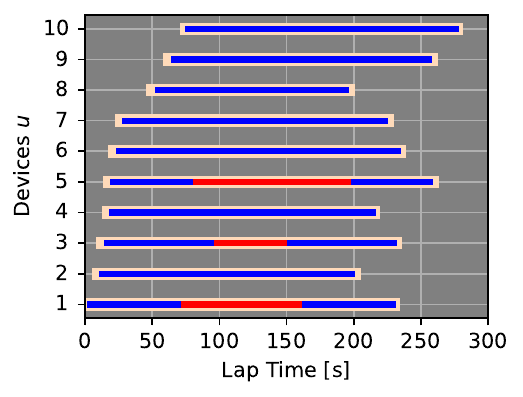}
         \vspace{-0.5cm} 
         \caption*{\hspace{0.2cm} (a) \hspace{9cm} (b)}
     \end{subfigure}
     \captionsetup{labelsep=period, name=Fig.}
     \caption{Visibility windows from device $u=1$ to device $u=10$ in peach color tone, and power levels: $\mathcal{L}_1$ (in blue) and $\mathcal{L}_2$ (in red). (a) In FTP, the devices $u \in \{1,3,5\}$ could generate both power levels in different opportunities while the other devices could generate only $\mathcal{L}_1$. (b) In CTP, nearly the entire visibility window can be exploited by the devices. }
    \label{fig:snapshot_scheme}
\end{figure*}

\begin{figure}[!t]
    \centering
    \includegraphics[width=\columnwidth]{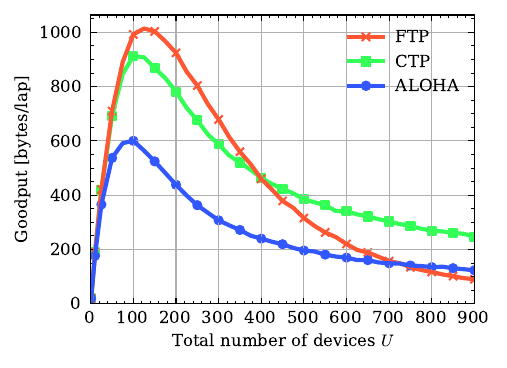} 
    \captionsetup{labelsep=period, name=Fig.}
    \caption{The average number of successfully received bytes per lap as a function of the number of IoT devices for the proposed FTP (in red) and CTP (in green) schemes, as well as for regular ALOHA (in blue).} 
    \label{fig:Goodput_MAC_DtS_schemes}
\end{figure}

\begin{figure}[!t]
    \centering
    \includegraphics[width=\columnwidth]{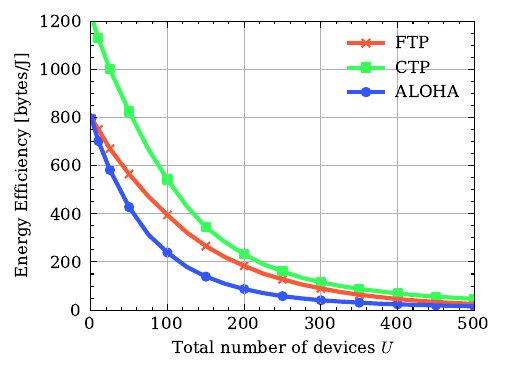} 
    \captionsetup{labelsep=period, name=Fig.}
    \caption{The energy efficiency as a function of the number of IoT devices for the proposed FTP (in red) and CTP (in green) schemes, as well as for regular ALOHA (in blue).} 
    \label{fig:Energy_Efficiency}
\end{figure}

Finally, although the methods proposed here can be applied to any number of power levels, in this work we constrain the application of SIC \felipe{to} only two power levels due to practical reasons. With more SIC rounds\felipe{,} the complexity at the gateway increases, the potential accumulation of residual interference can decrease the performance of SIC, and the probability to meet conditions ${\cal C}1$ and ${\cal C}$2 for several SIC iterations would certainly decrease considerably.

\section{Numerical Results} \label{sec:results}

\begin{table}[!t]
\caption{\justifying Simulation Parameters.}
\centering
\begin{tabular}{lrc}
\hline
\multicolumn{1}{c}{} & \multicolumn{1}{c}{Value} & \multicolumn{1}{l}{Parameter}     \\ \hline
Maximum Transmit Power             & $14$ {[}dBm{]}          &      $P$                       \\
Transmitter Antenna Gain \hspace{0.5cm}      & 0 {[}dBi{]} &    $G_{t}$                    \\
Receiver Antenna Gain         & $13.5$ {[}dBi{]}           &    $G_{r}$              \\
Channel Bandwidth              & $125$ {[}kHz{]}            &   $B$                           \\
Carrier Frequency      & $868$ {[}MHz{]}           &      $f$                        \\
Time-on-Air                    & $1.8104$ {[}s{]}           &      ToA                        \\
Payload size                   & $20$ {[}Bytes{]}           &      $b$                        \\
Spreading Factor                  & $12$                       &         SF                     \\
Sensitivity                    & $-137$ {[}dBm{]}           &         -                \\
SNR Threshold                  & $-20$ {[}dB{]}             &      $\gamma$                 \\
\felipe{SIR} Threshold         & $1$ {[}dB{]}               &     $\psi$                 \\ 
Noise Figure & 6 {[}dB{]} & $F$ \\ \hline
\end{tabular}
\label{tab:simulation_Parameters}
\end{table}

In this section\felipe{,} we present numerical results to evaluate the trade-off between the average number of successfully decoded transmissions and energy consumption considering the regular ALOHA protocol and the proposed schemes. The simulation parameters are listed in Table~\ref{tab:simulation_Parameters}. Aiming at a realistic scenario, the device locations are uniformly distributed over France according to geographic coordinates using the Python GeoPy library~\cite{GeoPy}. The distance between each device and the gateway is estimated with the Skyfield astronomy library~\cite{Skyfield}. Utilizing data from the CelesTrack platform~\cite{Celestrack}, this public library fits information in the two-line element (TLE) set format to determine the satellite's locations based on its orbit and pointing time. More specifically, the satellite visibility times allow us to compute the distance to the devices using~\cite{blufftonSkyfield}.  Furthermore, we consider the real orbit of the LacunaSat-3 LEO satellite, positioned at an altitude ranging from $500$~km to $600$~km above Earth~\cite{LS3.2020}. As in~\cite{Alvarez.22,Alonso.23}, we also assume a minimum elevation angle of $30^{\circ}$. Moreover, following~\cite{Afhamisis:ACS:2022, Alonso.23, F_Tondo.ASC.24, Ortigueira.21}, we consider the use of the most robust spreading factor.

We start by illustrating the behavior of the proposed \felipe{multiple-access} strategies. \felipe{Fig.~}\ref{fig:snapshot_scheme} shows a vulnerability analysis for $U=10$ devices considering $L=2$ power levels. In both cases, the peach color tone represents the visibility windows at the point of view of each device. Additionally, blue and red colors represent the time intervals during which the device is able to generate power levels ${\cal L}_1$ and ${\cal L}_2$  at the satellite, respectively. \felipe{Fig.~}\ref{fig:snapshot_scheme}a shows that, considering the FTP scheme, there are very specific intervals of time that $\mathcal{L}_1$ or $\mathcal{L}_2$ could be received at the LEO gateway, depending on the distance $d^{\text{FTP}}$. Moreover, in the case of CTP shown in \felipe{Fig.~}\ref{fig:snapshot_scheme}b, \felipe{the} IoT devices can more effectively utilize the  time window. In other words, selecting the required transmit power $P^{\text{CTP}}$\felipe{,} they can generate $\mathcal{L}_1$ or $\mathcal{L}_2$ at the satellite while transmitting at different positions within the visibility time window, spreading their transmissions. Finally, note that only devices \( u \in \{ 1, 3, 5 \} \) can generate two power levels, consequently, the FTP scheme allows for just four transmission opportunities, while CTP offers multiple opportunities for the uplink. Note that, in CTP, if the device chooses to generate power level ${\cal L}_1$\felipe{,} then it may transmit basically at anytime within the visibility window, but if it chooses ${\cal L}_2$\felipe{,} then such interval considerably decreases. This makes CTP a midday between FTP (very localized transmissions) and ALOHA (transmissions at anytime).

Next, \felipe{Fig.~}\ref{fig:Goodput_MAC_DtS_schemes} shows the goodput of the proposed DtS-IoT schemes, FTP (in red) and CTP (in green), as well as that of regular ALOHA (in blue) versus the number of devices and a payload of $b=20$ bytes. For $U=100$ devices\felipe{,} FTP achieves an improvement of \scd{$65\%$ ($992$ bytes/lap)} over regular ALOHA \scd{($601$ bytes/lap)}, while CTP performs \scd{$52\%$ ($913$ bytes/lap)} better than ALOHA. Note that, as illustrated by \felipe{Fig.~}\ref{fig:snapshot_scheme}, with a relatively small number of devices FTP has the advantage that collisions are less frequent since the devices transmit only at particular intervals in the visibility windows. However, as more devices appear in the satellite footprint, a large number of them have similar positions, leading to similar transmit time within their visibility windows. This issue increases the probability of collisions in FTP, so that CTP performs better for a sufficient number of devices. In the case of \felipe{Fig.~}\ref{fig:Goodput_MAC_DtS_schemes}, this happens for $U>400$. For instance, in the case of $U=600$, CTP and FTP achieve an improvement of \scd{$101\%$ ($341$ bytes/lap)} and \scd{$29\%$ ($220$ bytes/lap)} over regular ALOHA \scd{($170$ bytes/lap)}.

\scd{To evaluate our proposed schemes in terms of energy efficiency,} \felipe{Fig.~}\ref{fig:Energy_Efficiency} shows this metric versus the number of IoT devices $U$. Note that the CTP method offers a significant advantage in terms of energy efficiency with respect to FTP and ALOHA. With $U=100$ devices, CTP provides a gain of \scd{37\%} in terms of energy efficiency with respect to FTP, while this gain is of \scd{127\%} with respect to ALOHA. Moreover, CTP outperforms FTP and ALOHA for all numbers of IoT devices, compensating the fact that it is outperformed by FTP in terms of goodput for less dense networks. The advantage of CTP in terms of energy efficiency comes from the fact that the \felipe{devices can} adapt \felipe{their} transmit power (respecting a maximum transmit power constraint), leading to energy savings, what is very desirable in IoT networks.

\begin{figure}[t]
    \centering
    \includegraphics[width=\columnwidth]{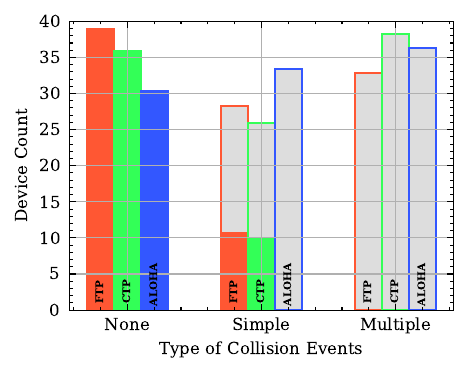} 
    \captionsetup{labelsep=period, name=Fig.}
    \caption{Device count \textit{versus} type of collision events considering  $U=100$ devices for FTP (in red), CTP (in green) and ALOHA (in blue). The collisions are classified as none, simple and multiple.} 
    \label{fig:Collision_U100}
\end{figure}

\begin{figure}[t]
    \centering
    \includegraphics[width=\columnwidth]{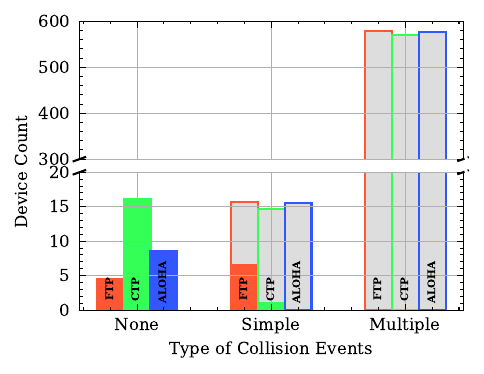} 
    \captionsetup{labelsep=period, name=Fig.}
    \caption{Device count \textit{versus} type of collision events considering  $U=600$ devices for FTP (in red), CTP (in green) and ALOHA (in blue). The collisions are classified as none, simple and multiple.} 
    \label{fig:Collision_U600}
\end{figure}

In order to better understand the goodput results, we \felipe{classify} the collision events for $U=100$ devices in \felipe{\felipe{Fig.~}\ref{fig:Collision_U100} as}:  (i) None, where the message was received without collisions; (ii) Simple, where a collision was detected between two transmitted messages; and (iii) Multiple, where more than two messages were received simultaneously. The number of collision events in each of the above three classes are shown for  FTP (bars in red), CTP (bars in green) and ALOHA (bars in blue). Moreover, the fraction of each bar filled in gray represents the collided messages that could not be decoded. For simple events, the satellite gateway can decode FTP and CTP messages if, and only if, they have different power levels (\textit{i.e}, one message was received with power $\mathcal{L}_1$ and the other with $\mathcal{L}_2$). \felipe{Fig.~}\ref{fig:Collision_U100} shows that FTP has   \scd{38.92} events of the type ``none'' per lap in average, more than the other methods, but also has more successfully decoded messages with simple collisions (\scd{10.68} events per lap in average). Consequently, FTP achieves the largest goodput (\scd{38.92 + 10.68} successfully decoded messages of 20 bytes each, leading to \scd{992} bytes/lap) among all for $U=100$ devices. Moreover,  \felipe{Fig.~}\ref{fig:Collision_U600} shows the same illustration but for $U=600$ devices, where the number of multiple collided messages considerably increases for all methods. Moreover, CTP presents many more events of the type ``none'' than FTP and ALOHA, leading to a larger goodput. In denser networks\felipe{,} it is more relevant to distribute the transmissions in larger time intervals as CTP does, what can be seen as a midway between what ALOHA does (the device is free to transmit at anytime within the visibility window) and what FTP does (the devices are allowed to transmit at only very particular points in the trajectory).

\section{Conclusion} \label{sec:conclusion}

This work presented two novel DtS-IoT \scd{multiple access} schemes using power domain NOMA. To explore particular positions of each IoT device within the visibility time window, we propose power domain NOMA strategies, FTP and CTP, using either fixed and controlled transmit power. We evaluated the goodput and energy efficiency of both strategies and compared with regular ALOHA. The proposed methods greatly outperform ALOHA in terms of goodput, with FTP performing better up to a number of devices, while CTP performs better for denser networks. Moreover, in terms of energy efficiency CTP showed to be superior for all numbers of devices. \scd{In future works, we intend to consider device traffic characteristics to enhance the proposed approaches. While energy efficiency has been evaluated in this work, we aim to explore the flexibility of our proposed protocols in adapting to energy constraints and awareness (typical of IoT setups), including energy harvesting possibilities.} 

\input{main_V1.bbl}

\end{document}